\documentclass{emulateapj}

\begin{document}

\title{The Impact of ICM Substructure on Ram Pressure Stripping}
\author{Stephanie Tonnesen and Greg L. Bryan}
\affil{Department of Astronomy, Columbia University, Pupin Physics Laboratories, New York, NY 10027}

\begin{abstract}

Cluster galaxies moving through the intracluster medium (ICM) are expected to lose some of their interstellar medium (ISM) through ram pressure stripping and related ISM-ICM interactions.  Using high-resolution cosmological simulations of a large galaxy cluster including star formation, we show that the ram pressure a galaxy experiences at a fixed distance from the cluster center can vary by well over an order of magnitude  We find that this variation in ram pressure is due in almost equal parts to variation in the ICM density and in the relative velocity between the galaxy and the ICM.  We also find that the ICM and galaxy velocities are weakly correlated for in-falling galaxies.

\end{abstract}

\keywords{galaxies: clusters, galaxies: interactions, methods: N-body simulations}

\section{Introduction}

X-ray observations of clusters have shown that substructure in the intracluster medium (ICM) is common (e.g. Mohr, Mathiesen, \& Evrard 1999; Schuecker et al 2001).  In a sample of 470 clusters, Schuecker et al. (2001) measure substructure in more than 50{\%} of their sample.  Detailed examinations of nearby clusters like Perseus and Virgo have discovered substructure and/or asymmetry in both the temperature and density profiles of these clusters (e.g. Bohringer et al. 1994; Shibata et al. 2001; Churazov et al. 2003; Dupke \& Bregman 2001; Furusho et al 2001). Even Coma, considered a relaxed cluster, has ICM irregularities (White, Briel \& Henry 1993).  The importance of substructure on cluster mass measurements has been examined (Mohr, Mathiesen \& Evrard 1999; Bohringer et al 2000), which in turn affects the use of cluster measurements as cosmological constraints (Jeltema et al. 2005; Nagai et al. 2007).

However, the importance of substructure in the ICM is rarely considered when studying ram pressure stripped galaxies.  Common assumptions are that the ICM is static, has a smooth density profile, and is only dense enough very near the center of a cluster to affect galaxies.  Treu et al. (2003), in their evaluation of possible environmental evolutionary mechanisms in Cl 0024 + 16, assume that ram pressure is only effective to 0.6 virial radii.  Solanes et al. (2001) find HI deficiency in galaxies out to two Abell radii, but only discuss the possibility that these galaxies are on highly radial orbits that have already carried them through the cluster center.  Previous simulations studying galaxy evolution in clusters use a static, smooth ICM profile when studying the orbits of galaxies in clusters (e.g. Vollmer 2001; Roediger \& Br{\"u}ggen 2007; J{\'a}chym et al. 2007).  These authors use different galaxy orbits in order to sample a variety of galaxy velocities at a fixed ICM density. 

Although the use of simple assumptions is widespread, there is at least one possible case in which ICM substructure had to be invoked to explain observations of the Virgo galaxy NGC 4522, a galaxy with a truncated gas disk (Kenney et al 2004; Vollmer et al. 2004; Vollmer et al. 2006).  NGC 4522 is located at a projected distance of 1 Mpc from the center of the Virgo cluster, and assuming a static ICM with standard density values, the ram pressure is not strong enough to cause the observed truncation.  Thus, the authors propose that the nearby ICM is either moving relative to the galaxy or overdense.  

In a recent paper studying the environmentally-driven evolution of galaxies in clusters using a detailed cosmological simulation (Tonnesen, Bryan \& van Gorkom 2007), we examined the evolution of cool gas (i.e. ISM) in galaxies within and around the cluster, demonstrating that most gas loss from galaxies was due to ISM-ICM interactions (i.e. ram pressure and related processes), rather than galaxy-galaxy interactions or cluster tidal effects.  We also found that ram pressure stripping occurs out to the virial radius of the cluster (measured using $r_{200}$). 

In this paper, we examine this result more closely and show that the ram pressure a galaxy experiences varies substantially, even at fixed distance from the cluster center.   As we will see, this arises both from the density and velocity substructure of the ICM.   First, we briefly introduce our code in \S \ref{sec:sim} and explain how we measure ram pressure in our simulation (\S \ref {sec:sample} and \S \ref{sec:rpsample}).  We then present our results:  a comparison of the standard deviations of ram pressure, ICM density, and velocity difference squared (\S \ref{sec:density}), followed by a more detailed look at the velocity of the ICM (\S \ref{sec:velocity}).  

\section{Methodology}

\subsection{Simulation}
\label{sec:sim}

We have simulated a massive cluster of galaxies with the adaptive mesh refinement (AMR) code {\it Enzo}.  This cosmological hydrodynamics code uses particles to evolve the dark matter and stellar components, while using an adaptive mesh for solving the fluid equations including gravity (Bryan 1999; Norman \& Bryan 1999; O'Shea et al. 2004).  The code begins with a fixed, static grid and automatically adds refined grids as required in order to resolve important features in the flow (as defined by enhanced density).  An image of this cluster is shown in Figure 1 of our earlier paper (Tonnesen et al. 2007), and visualizations of these simulations can be found at http://www.astro.columbia.edu/\~{}gbryan/ClusterMovies.

We chose to examine the largest  cluster ($r_{200}$ is 1.8 Mpc and $M_{200}$ is $6 \times 10^{14}$ M$_\odot$) that formed within a periodic simulation box which was 64 $h^{-1}$ Mpc on a side, in a flat, cosmological-constant dominated universe with the following parameters: $(\Omega_0, \Omega_\Lambda, \Omega_b, h, \sigma_8) = (0.3,0.7, 0.045, 0.7, 0.9)$.  We employ a multi-mass initialization technique in order to provide high-resolution in the region surrounding the cluster, while evolving the rest of the box at low resolution.  The dark-matter particle mass is $6.4 \times 10^{8}$ M$_\odot$, with a gas mass resolution about five times better than this. The whole cluster has more than one million particles within the virial radius, and a typical $L_*$ galaxy is resolved by several thousand particles.  The adaptive mesh refinement provides higher resolution in high density regions, giving a best cell size (resolution) of 3 kpc.

The simulation includes radiative cooling using the White \& Sarazin (1987) cooling curve, and an approximate form of star formation and supernovae feedback following the Cen \& Ostriker (1992) model.  More details can be found in our earlier paper (Tonnesen et al. 2007).

%Fig 1
\begin{figure*}
\includegraphics{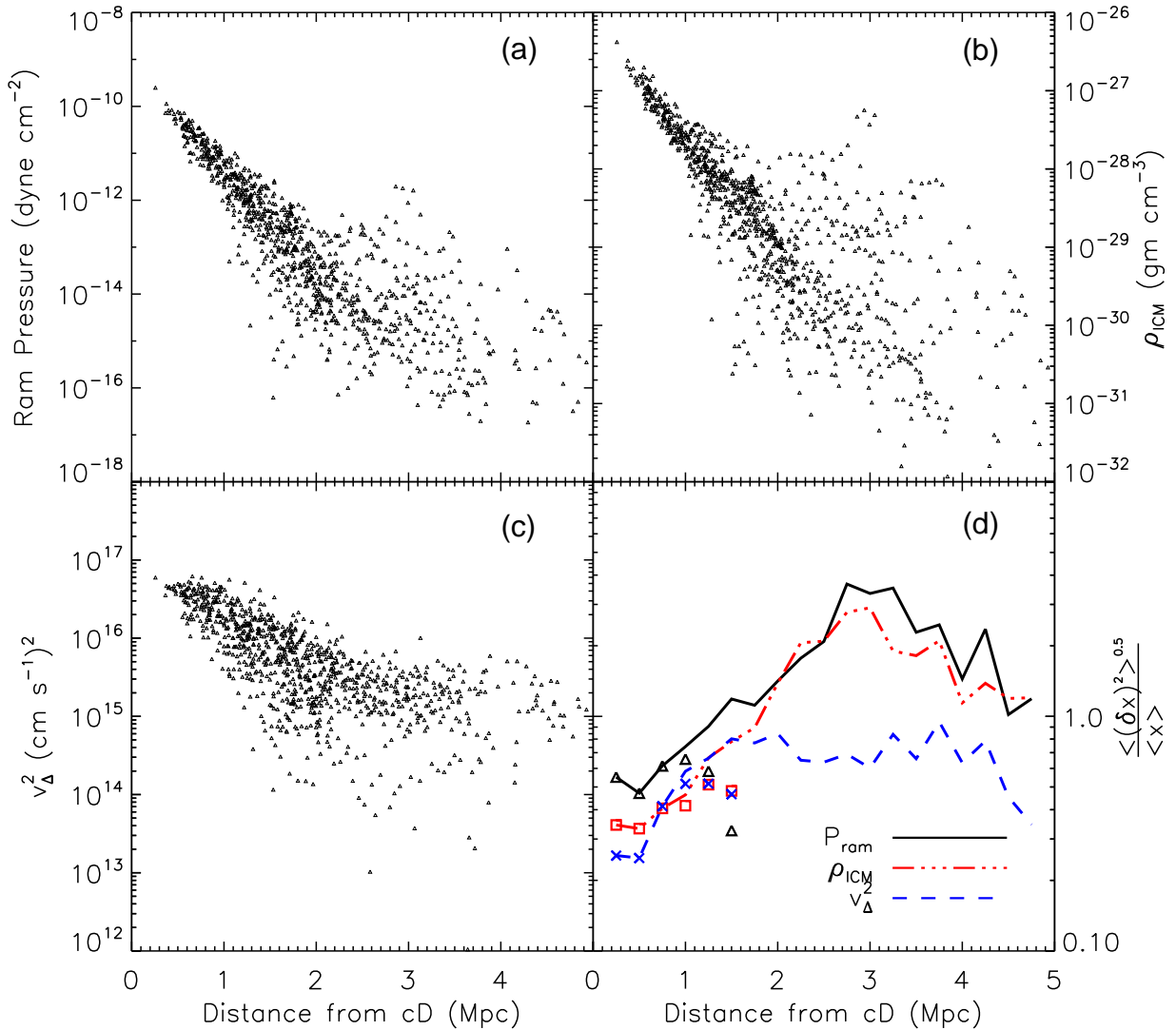}
\caption{This figure shows the variation in (a) ram pressure, (b) ICM density and (c) the square of the velocity difference between the ICM and galaxies ($v_\Delta^2$), all as a function of distance from the cD.  In panel (d) lines are the normalized standard deviation as a function of distance from the cD for each of these three measurements, while the symbols show the same quantity for galaxies experiencing ram pressure greater than $10^{-12}$ dynes cm$^{-2}$ (triangles are ram pressure, squares are ICM density, and X's are $v^2_{\Delta}$).  See \S \ref{sec:density} for discussion.}
\label{fig:4panel}
\end{figure*}

\subsection{Construction of our Sample}
\label {sec:sample}

In order to accurately determine the ICM conditions a gas-rich galaxy experiences as it falls into the cluster, we identify and track a sample of galaxies which form in the simulation.  This naturally gives us realistic galactic trajectories.  In order to construct the sample, we first separate our star particles into distinct galaxies based on regions of high-density in our N-body stellar code.  A visual inspection of the data shows that, as in real clusters, galaxies are easy to identify because they are highly concentrated, with relatively few stars between galaxies.  

We used the HOP algorithm (Eisenstein {\&} Hut, 1998), which uses a two-step procedure to identify individual galaxies.  First, the algorithm assigns a density to each star particle based on the distribution of the surrounding particles and then hops from a particle to its densest nearby neighbor until a maximum is reached. All particles (with densities above a minimum threshold, $\delta_{\rm outer}$) that reach the same maximum are identified as one coherent group.  In the second step, groups are combined if the density at the saddle point which connects them is greater than $\delta_{\rm saddle}$.  We use HOP because of its physical basis, although we expect similar results would be found using a friends-of-friends halo finder.  We identify all such galaxies in $33$ outputs over 3.5 Gyr and then form trajectories by identifying the same galaxy in all outputs.  The galaxies identified and followed are most often near or above the mass of the Milky Way, although we do follow a few that are about a third of the Milky Way galaxy's mass.  For more details, see our earlier paper (Tonnesen et al. 2007). 

\subsection{Measuring Ram Pressure}
\label{sec:rpsample}

After identifying all of the galaxies in our cluster, we measure ram pressure only around galaxies that have cool gas (T $\le$ 15,000 K), as these are the galaxies it will affect.  Ram pressure is measured by using the Gunn \& Gott (1972) equation $P_{ram} = \rho v^{2}_\Delta$, where $\rho$ is the ICM density and $v_\Delta$ is the velocity difference between the ICM and the galaxy ($|\vec{v}_{galaxy} - \vec{v}_{ICM}|$).  For the galaxy velocity, we adopt the mean of the the cool gas within a 26.7 kpc sphere around the center of the galaxy.  As described in detail in our earlier paper (Tonnesen et al. 2007), we chose this radius because it excluded gas from nearby galaxies while containing the gas from the galaxy we followed.  Gas was defined to be part of the ICM if it had a temperature above $10^7$ K.  A galaxy's local ICM properties were determined by averaging the density and velocity of all ICM gas within a 90 kpc sphere centered on each galaxy.  We compared these ICM measurements to ones taken using a mass-weighted average and an average only in an annulus from 26.7 kpc to 90 kpc, finding no qualitative difference and negligible quantitative difference in our results.

\section{Results}

\subsection{Effect of ICM Substructure on Ram Pressure}
\label{sec:density}

In Figure \ref{fig:4panel} (a), we show the ram pressure experienced by galaxies as a function of distance from the cluster center.  Although there is a strong trend of decreasing ram pressure with cluster distance, there is also a substantial scatter at fixed radius.  
%For example, at 1 Mpc from the cD galaxy, the central 80\% of measured ram pressure values span an order of magnitude, while at  the virial radius, 1.8 Mpc, the values span almost two orders of magnitude.  
This suggests that the assumptions about the ICM used by observers and theorists alike to understand galaxy-ICM interactions may be too simplified.

To explore the origin of this scatter, we plot in Figures \ref{fig:4panel} (b) and (c) the ICM density and $v^2_{\Delta}$ as a function of radius.  From these figures, we see that at 1 Mpc cluster radius, the central 80\% of ram pressure values range over an order of magnitude, while the ICM density and square of the velocity vary by factors of three and six, respectively.  At the virial radius (1.8 Mpc) the ram pressure varies across almost two orders of magnitude and both the ICM density and $v^2_{\Delta}$ vary by at least an order of magnitude.

In order to make this more quantitative, we measure the variance of all three values in radial bins of 250 kpc width, normalizing the standard deviation by the mean of the value measured in each bin.  Bin size does not affect our conclusions.  Our results are shown in Figure \ref{fig:4panel} (d).  In this figure the normalized standard deviation of the ram pressure is the solid black line, the ICM density is the dash-dotted red line, and $v^2_\Delta$ is the dashed blue line.   We note that the sum of the variances of the two components (i.e. density and $v^2_\Delta$) closely matches the variance in the ram pressure, indicating that the two components are uncorrelated.

Within the virial radius of the cluster, the standard deviation of the ICM density and $v^2_\Delta$ are very similar.  The inner region is where most ram pressure stripping is thought to take place; we also found the most ram pressure stripping in our simulated cluster within the virial radius.  Even in this region, different orbits of galaxies, and the resulting different galaxy velocities at a fixed cluster radius, are no more important than ICM density fluctuations in determining ram pressure.  Outside the virial radius of the cluster, the standard deviation of the ICM density is higher than that of $v^2_\Delta$, and more influential on the variation of the ram pressure.  Far from the cluster core (r $\gg 4$ Mpc), we see that galaxies falling into the cluster for the first time are distributed in regions of both high and low density.  Visualizations of this simulation show that at low redshift most galaxies fall into the cluster along a wide filament, which must also have a large scatter in density (see Dav{\'e} et al. (2001) for a detailed discussion of the density of the Warm Hot Ionized Medium).

Recall that this cluster has structure not only in the ICM density, but also in the ICM velocity.  To check that the ICM velocity is not reducing the standard deviation of $v^2_\Delta$, we also extracted the normalized standard deviation of $v^2_{galaxy}$.  This would be $v^2_\Delta$ if the ICM velocity were zero.  The scatter of this value is even smaller than the scatter of $v^2_\Delta$, so in a static ICM the density substructure would be even more important in varying ram pressure at fixed cluster radius.

The lines in Figure \ref{fig:4panel} (d) include all of our data points, including ram pressure values well below those that could strip a galaxy of its gas.  To check that this does not effect our conclusion, we also plot the standard deviation for the three variables using only ram pressure values greater than $10^{-12}$ dynes cm$^{-2}$, the Gunn \& Gott (1972) limit for ram pressure stripping to effect a Milky Way sized galaxy, as symbols with the same color scheme as the lines.  Note that when we include only these values, the ram pressure variation is still contributed equally by density and velocity variations.  
%We believe that the values measured at 1.5 Mpc are influenced by small number statistics.  
It is clear from these results, whether we include all of our data or only those points with high ram pressure, that it is unrealistic to assume that the scatter in ram pressure values at a fixed radius arises mainly from varying galaxy velocities using different orbits.

%Fig 2
\begin{figure}
\includegraphics[scale=0.45]{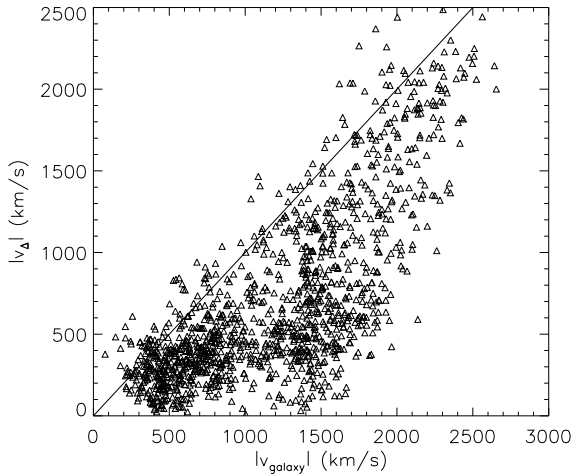}
\caption{ The correlation between $v_\Delta$ (which is $v_{galaxy} - v_{ICM}$) and the galaxy velocity for galaxies with gas.  If the ICM were static, all points would lie along the solid line.  Most points lie below this line, indicating that ICM and galaxy velocities are correlated. See \S \ref{sec:velocity} for details.
}
\label{fig:v_compare}
\end{figure}

\subsection{ICM Velocity Structure}
\label{sec:velocity}

Since we track galaxies moving through the ICM, we can also critically examine the motions of the ICM gas that these galaxies experience.  We find that ICM velocity is correlated with galaxy velocity, and therefore our measured ram pressure is smaller than would be found if we assumed a static ICM.   We have plotted the magnitude of $v_\Delta$ against the magnitude of the galaxy velocity in Figure \ref{fig:v_compare}.  To guide the eye, we have drawn a line of equality, on which the points would fall if the ICM were static.  The vast majority of the velocity difference measurements, particularly for low galaxy velocities, are smaller than the galaxy velocity.  Recall that we are only following galaxies that have cool ($\le$ 15,000 K) gas, which are dominated by galaxies falling towards the cluster center.  The ICM velocity and galaxy velocity are correlated because the ICM is also falling towards the cluster center.  This is true throughout the 3.5 Gyr adopted for our analysis, during which time no major merger event occurs that would re-disturb the ICM.  From the visualizations of these simulations, the last significant merger occurred at a redshift of about 0.5.  The sound crossing time at the virial radius (using an ICM temperature of $4 \times 10^7$ K) is less than 2 Gyr, so in simple models the ICM would equilibrate within the time we study the cluster.  Again, it is clear that the most simple assumption cannot well describe the ICM or its impact on ram pressure stripping, nor does the static assumption result in a median of the measured values.

\section{Conclusion}
\label{sec:discussion}

In this paper we have presented a detailed examination of the intracluster medium with which a galaxy interacts as it falls into a simulated galaxy cluster.  We find that substructure in the ICM is more important in varying ram pressure than is often assumed and used when modeling ram pressure stripping.  Specifically, we highlight three main points:

\begin{enumerate}

\item  In our simulated cluster we measure a range of ram pressure values for any given radius in the cluster.  This ranges from an order of magnitude at 1 Mpc, to two orders of magnitude at the virial radius (1.8 Mpc), to even larger deviations further from the cD.  Therefore, ram pressure can be effective at larger radii wherever there is an overdensity.

\item  The scatter in ram pressure at different distances from the cD is due equally to the variation in the ICM density and the relative ICM-galaxy velocity ($v_{\Delta}^2$) within the virial radius.  This is true even when considering only higher ram pressure values.  In fact, the normalized standard deviation in galaxy velocity is smaller than that of $v_{\Delta}^2$.  It is therefore not only the variety of orbital velocities that causes different values of ram pressure at fixed cluster radius, but also the density and velocity structure of the ICM.  Further from the cD, ICM density variations dominate those of $v_{\Delta}^2$.  

\item  The ICM velocity is correlated with galaxy velocity, resulting in a smaller $v_{\Delta}$ than $v_{galaxy}$.  This indicates that the ICM tends to move with in-falling galaxies, which then experience somewhat less ram-pressure than one would expect from a static ICM (although this is less true for high-velocity galaxies that are likely near the cluster center).

\end{enumerate}

We emphasize that although we determine the ICM properties from a simulation, it is well-known in the X-ray cluster field that ICM substructure in density is common and in good agreement with simulations (Mohr, Mathiesen, \& Evrard 1999; Jeltema et al. 2005; Nagai et al. 2007).  Because ram pressure stripping is a fast process, even an overdensity with a relatively small extent can strip a galaxy that might otherwise retain its gas, or strip a galaxy more than predicted by its cluster position.  Our results should galvanize the community currently studying galaxy evolution in clusters to look more closely at the intracluster medium.

\acknowledgements

GB acknowledges support from NSF grants AST-05-07161, AST-05-47823, and
AST-06-06959, as well as computational resources from the National
Center for Supercomputing Applications.  We thank Jacqueline van Gorkom for
her help with this paper.

\end{document}